\documentclass[pdflatex, sn-mathphys-num]{sn-jnl}

\usepackage{graphicx}%
\usepackage{multirow}%
\usepackage{tabularx}
\usepackage{amsmath,amssymb,amsfonts}%
\usepackage{amsthm}%
\usepackage{mathrsfs}%
\usepackage[title]{appendix}%
\usepackage{xcolor}%
\usepackage{textcomp}%
\usepackage{manyfoot}%
\usepackage{booktabs}%
\usepackage{algorithm}%
\usepackage{algorithmicx}%
\usepackage{algpseudocode}%
\usepackage{listings}%
\usepackage{array}
\usepackage[tableposition=top]{caption}
\usepackage{orcidlink}
\usepackage{lmodern}

\usepackage{float}
\floatstyle{plaintop}
\restylefloat{table}

\begin{document}

\title[Article Title]{\textit{KEVS}: Enhancing Segmentation of Visceral Adipose Tissue in Pre-Cystectomy CT with Gaussian Kernel Density Estimation}


\author*[1,3]{\fnm{\orcidlink{0009-0002-6152-7271}Thomas} \sur{Boucher}}\email{\{thomas.boucher.23; e.mazomenos\}@ucl.ac.uk}
\author[2,3]{\fnm{\orcidlink{0000-0002-4907-9018}Nicholas} \sur{Tetlow}}
\author[2,3]{\fnm{\orcidlink{0000-0001-6688-904X}Annie} \sur{Fung}}
\author[2,3]{\fnm{\orcidlink{0009-0007-3015-7040}Amy} \sur{Dewar}}
\author[3]{\fnm{\orcidlink{0000-0002-2080-5274}Pietro} \sur{Arina}}
\author[2,3]{\fnm{\orcidlink{0000-0002-9683-2295}Sven} \sur{Kerneis}}
\author[2,3]{\fnm{\orcidlink{0000-0002-3859-679X}John} \sur{Whittle}}
\author*[1]{\fnm{\orcidlink{0000-0003-0357-5996}Evangelos B.} \sur{Mazomenos}}

\affil*[1]{\orgdiv{UCL Hawkes Institute, Department of Medical Physics and Biomedical Engineering}, \orgname{UCL}, \orgaddress{\city{London}, \country{UK}}}
\affil[2]{\orgdiv{Department of Anaesthesia and Peri-operative Medicine}, \orgname{University College London Hospitals NHS Foundation Trust}, \orgaddress{\city{London}, \country{UK}}}
\affil[3]{\orgdiv{Human Physiology and Performance Laboratory (HPPL),
Centre for Peri-operative Medicine, Department of Targeted Intervention, Division of Surgery and Interventional Science}, \orgname{UCL}, \orgaddress{\city{London}, \country{UK}}}
\abstract{\textbf{Purpose:} The distribution of visceral adipose tissue (VAT) in cystectomy patients is indicative of the incidence of post-operative complications. Existing VAT segmentation methods for computed tomography (CT) employing intensity thresholding have limitations relating to inter-observer variability. Moreover, the difficulty in creating ground-truth masks limits the development of deep learning (DL) models for this task. This paper introduces a novel method for VAT prediction in pre-cystectomy CT, which is fully automated and does not require ground-truth VAT masks for training, overcoming aforementioned limitations.

\textbf{Methods:} We introduce the \textit{Kernel density Enhanced VAT Segmentator (}\textit{KEVS}\textit{)}, combining a DL semantic segmentation model, for multi-body feature prediction, with Gaussian kernel density estimation analysis of predicted subcutaneous adipose tissue to achieve accurate scan-specific predictions of VAT in the abdominal cavity. Uniquely for a DL pipeline, \textit{KEVS} does not require ground-truth VAT masks. 

\textbf{Results:} We verify the ability of \textit{KEVS} to accurately segment abdominal organs in unseen CT data and compare \textit{KEVS} VAT segmentation predictions to existing state-of-the-art (SOTA) approaches in a dataset of 20 pre-cystectomy CT scans, collected from University College London Hospital (UCLH-Cyst), with expert ground-truth annotations. \textit{KEVS} presents a $4.80\%$ and $6.02\%$ improvement in Dice Coefficient over the second best DL and thresholding-based VAT segmentation techniques respectively when evaluated on UCLH-Cyst. 

\textbf{Conclusion:} This research introduces \textit{KEVS}; an automated, SOTA method for the prediction of VAT in pre-cystectomy CT which eliminates inter-observer variability and is trained entirely on open-source CT datasets which do not contain ground-truth VAT masks.}
%
%
%
\maketitle
\section{Introduction}\label{sec1}
Body composition is a key predictor of complications following cystectomy, the removal of the urinary bladder, usually for cancer, where abnormal adiposity is associated with worse perioperative outcomes and increased cost to healthcare providers \cite{huynh2020cost,deuker2021obesity}. To properly assess body composition, it is then necessary to accurately determine the distribution of adipose tissue (AT) in a patient. 
Two of the most significant deposits of AT in the body are subcutaneous adipose tissue (SAT), found just below the skin, and visceral adipose tissue (VAT), which surrounds the internal organs.
Peri-operative attention focuses predominately on abdominal VAT, as abnormal abdominal visceral adiposity is strongly linked with poor post-surgical outcomes \cite{tsujinaka2008visceral,watanabe2014impact}. 

State-of-the-art (SOTA) VAT prediction methods in CT rely either on pre-defined Hounsfield Unit (HU) thresholding on the abdominal region and more recently on deep learning (DL) models.
Segmentation with HU thresholding assigns voxels with intensity values within a pre-defined range to be VAT. This has notable limitations as there is often a lack of consensus about the most appropriate thresholding range to use \cite{noumura2017visceral}. Inevitably, this introduces inter-observer variability into VAT predictions, as they will depend on the chosen HU threshold. This subjectivity affects the consistency of segmentations between experts \cite{joskowicz2019inter,jungo2018effect}. Additionally, specific thresholds may not be appropriate for all types of CT scan which limits the generalisability and robustness of this approach. For example, low-dose CT scans tend to have wider distributions of HU compared to regular-dose CT due to additional noise \cite{padgett2014local}, therefore it would be inappropriate to use the same threshold on both types of scan.

Whilst deploying DL models for semantic segmentation in medical imaging has met great success \cite{zhou2021}, there are tangible limitations in using supervised learning for VAT segmentation due to the non-uniform distribution of this deposit in the abdominal cavity. In \cite{sundar2022fully}, the authors used HU thresholding to produce pseudo ground-truth masks for training an nnU-Net network \cite{isensee2021nnu}. This technique effectively introduces the limitations discussed previously to their model, and affects the validity of using the associated public dataset \cite{selfridge2023low} for training DL models. In \cite{lee2021deep}, the production of ground-truth VAT masks required the annotation of over 40,000 axial CT slices for training and validating $2D$ and $3D$ U-Nets, which is extremely time consuming, even with the semi-automated approach followed where HU thresholding produces initial masks, due to the non-uniformity of VAT. Neither the dataset nor the trained models from this work are made publicly available. 

This paper introduces the \textit{KDE-Enhanced VAT Segmentator (\textit{KEVS})}, a novel pipeline for fully automated volumetric prediction of VAT in CT. \textit{KEVS} combines semantic segmentation models, trained on the publicly available TotalSegmentator \cite{wasserthal2023totalsegmentator} and Sparsely Annotated Region and Organ Segmentation (SAROS) datasets \cite{koitka2024saros}, with a Gaussian Kernel Density Estimation (GKDE) fit to predicted SAT intensity values to ultimately segment VAT. 
%
%
The key innovation is that we can reliably identify candidate VAT voxels by accurately segmenting the abdominal cavity, abdominal organs, and SAT with a model trained on these two datasets. We can then exclude the organ predictions from the abdominal cavity, significantly reducing the number of voxels under consideration for VAT. Then, we assign candidate voxels as VAT according to their consistency with the intensity distribution, modelled via GKDE, of predicted SAT voxels.


Our research makes the following key contributions: \textbf{(1)} We introduce \textit{KEVS}, a novel method for the prediction of VAT in pre-cystectomy CT combining both semantic segmentation and GKDE; \textbf{(2)} \textit{KEVS} demonstrates SOTA performance with an average Dice Coefficient of $0.8697$ outperforming the best HU thresholding predictions ($0.8203$) and the open-source TotalSegmentator tissue segmentation model ($0.8298$), in a dataset of 20 pre-cystectomy scans recorded at University College London Hopsital (UCLH-Cyst); \textbf{(3)}\textit{KEVS} represents the first attempt to combine semantic segmentation models with scan-specific SAT intensity distribution analysis to estimate volumetric VAT in CT scans; \textbf{(4)} We provide UCLH-Cyst; a new dataset with VAT annotations in pre-cystecomy CT.

\begin{figure}[t]
    \centering
    \includegraphics[width=1\textwidth]{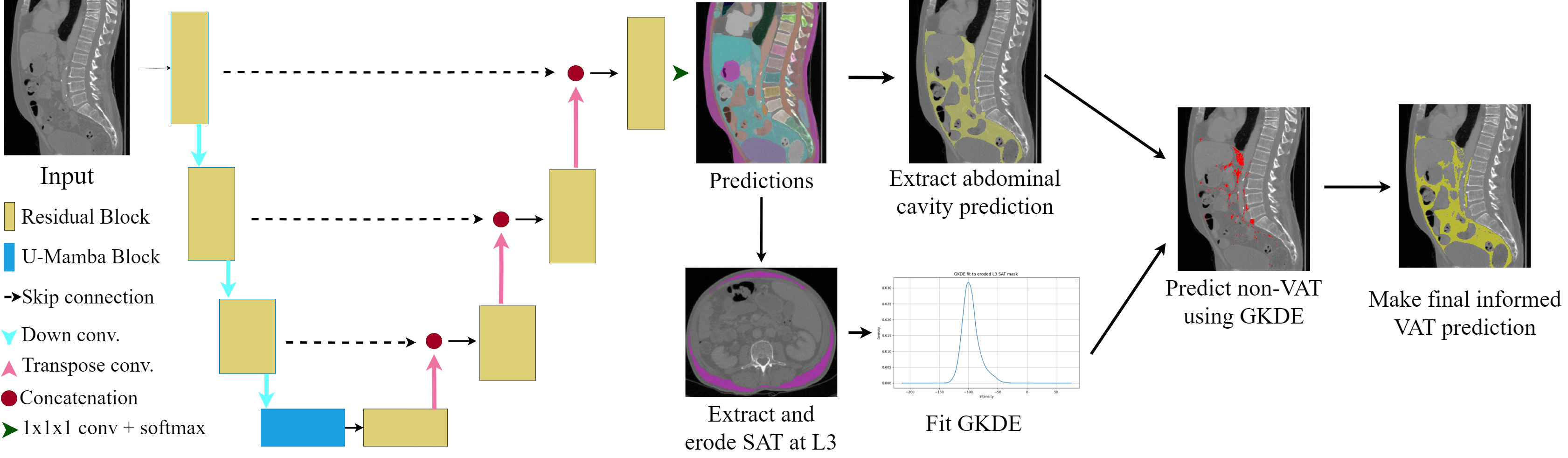}
    \caption{ A depiction of the full implementation of \textit{KEVS}. First, a U-Mamba architecture trained on our extended TotalSegmentator dataset makes a prediction on a CT scan. The abdominal cavity and \textit{L3} axial SAT prediction are extracted simultaneously, the SAT prediction is eroded and a GKDE fit to the resulting pixel intensities. This GKDE is used to predict the voxels in the abdominal cavity mask which are least likely to be VAT, which are removed from the abdominal cavity mask in the final step to give an informed prediction for VAT. }
    \label{fig:Fig1}
\end{figure}
\section{Methods}\label{sec2}
The overview of the \textit{KEVS} architecture is depicted in Fig.~\ref{fig:Fig1}. \textit{KEVS} achieves segmentation of VAT in $3D$ CT by combining DL based semantic segmentation and GKDE. For segmentation, we trained the nnU-Net based U-Mamba model, which improves representation of long-range dependencies over baseline nnU-Net \cite{gu2023mamba, ma2024u}.
Most crucial for our task is the ability to segment the abdominal cavity, abdominal organs, and SAT. The SAT predictions are analysed using GKDE, and the fitted kernel is applied to abdominal cavity voxels. Finally, the $15\%$ of abdominal cavity intensity values with the lowest probability density are removed from the abdominal cavity mask, and the remaining voxels are assigned as the final VAT segmentation. Our motivation for the GKDE analysis of the predicted SAT is to obtain an accurate continuous approximation of the distribution of the AT intensity values in an automated, thus without suffering from inter-observer variability, and scan-specific manner, thus directly applicable to any-dose CT scan. 
\subsection{Training Datasets}
To extract both the abdominal region and the abdominal organs with high accuracy we trained U-Mamba models on the SAROS and TotalSegmentator datasets:

\textbf{SAROS} \cite{koitka2024saros} provides annotations for every fifth axial slice in $3D$ whole-body and abdominal CT scans for 14 classes; 13 body features and the background. Crucially, this dataset contains ground-truth masks for the abdominal cavity and SAT. We extracted only the annotated slices from each patient (726 total patients for non-restricted datasets), and created a dataset of annotated $2D$ axial slices. 
We randomly chose a $626/100$, train/test patient split, where the training set was then partitioned into $501/125$, train/validation split. In total, $7917/1979/3722$ slices were used in the train/validation/test split.

\textbf{TotalSegmentator} \cite{wasserthal2023totalsegmentator} is comprised of $1228$ CT scans, each with ground-truth annotations for 118 different classes; 117 body features and the background. Crucially, this dataset contains ground-truth masks for the abdominal organs. We randomly selected $10\%$ as a testing set, accounting for $123$ scans. The remaining $1105$ scans are allocated into train/validation with an $80\%/20\%$ split, accounting for $884$ training and $221$ validation scans. 
\begin{figure}[h]
    \centering
    \includegraphics[width=1\textwidth]{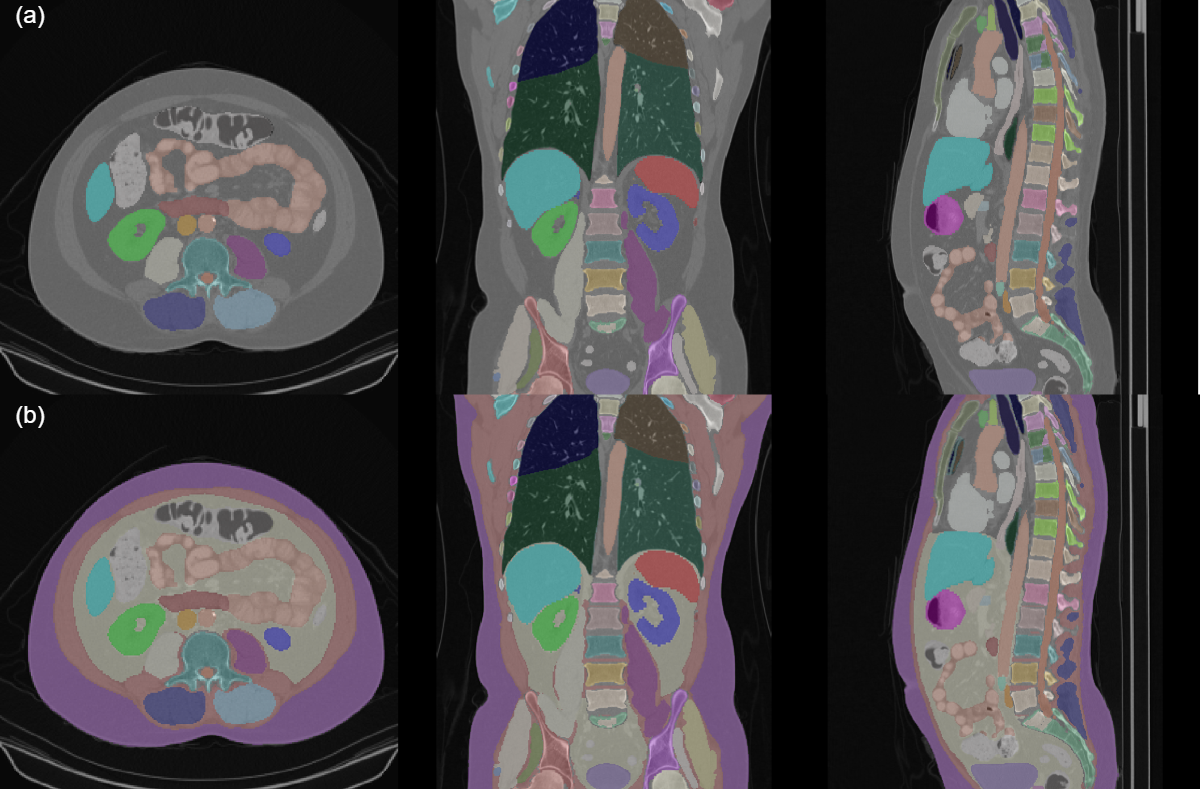}
    \caption{ \textbf{(a)} An example of the TotalSegmentator dataset for a coronal, axial, and sagittal view of a thoracic/abdominal CT scan. \textbf{(b)} An example of the combined SAROS predictions and TotalSegmentator ground-truth.}
    \label{fig:Fig2}
\end{figure}
\subsection{\textit{KEVS} Implementation}\label{subsection2.2}
\begin{figure}[t]
   \centering
    \includegraphics[width=1\textwidth]{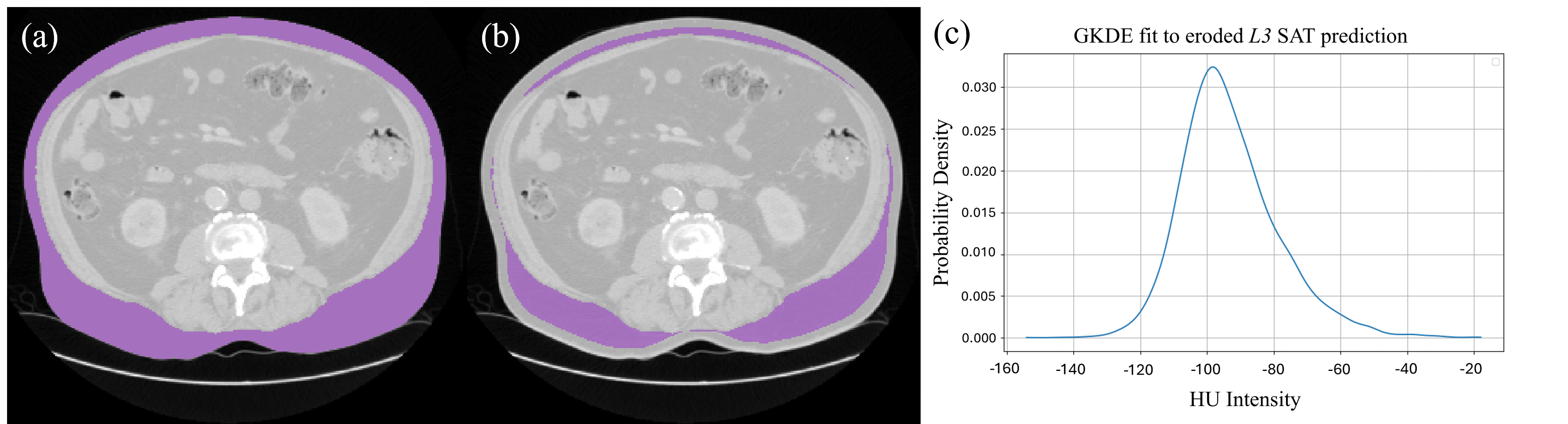}
    \caption{ \textbf{(a)}  An axial CT slice at the predicted \textit{L3} level with predicted SAT (purple) mask. \textbf{(b)} The same CT with the SAT mask eroded to $20\%$ of its original area. \textbf{(c)} A visualisation of the GKDE fit to the eroded SAT mask.}
\label{fig:Fig3}
\end{figure}

\textbf{Semantic Segmentation}\\
We employed the recently proposed nnU-Net based U-Mamba architecture due to its SOTA performance for medical semantic segmentation on a variety of datasets \cite{ma2024u, liu2024swin}. 
%
%
The U-Mamba block contains two successive residual blocks followed by a Mamba block for long-range dependency modelling. We selected the \textit{U-Mamba\_bot} variant where the U-Mamba block is built only into the bottle neck as it is less computationally intensive without compromising performance. 
We initially trained a U-Mamba model on the SAROS dataset to make predictions on the TotalSegmentator dataset. These predictions were merged with the TotalSegmentator ground-truth masks, prioritising the ground-truth where they overlap, to build a richly annotated dataset incorporating SAT and abdominal cavity pseudo ground-truth, for training a single model. Fig.~\ref{fig:Fig2} shows an example of this process. 
We then trained a U-Mamba model on the combined dataset. As shown in  Fig.~\ref{fig:Fig1}, \textit{KEVS} then accurately segments both SAT and an organ-free abdominal cavity.
\\
Networks were trained using combined cross-entropy and Dice loss with equal weighting, a Stochastic Gradient Decent optimiser, weight decay of $3\times10^{-5}$, and initial learning rate of $1\times10^{-2}$, which decreased to $1\times10^{-5}$ using a polynomial scheduler according to the formula $\text{\textit{lr}} = (1\times10^{-2})\cdot(1-\frac{epoch-1}{N})^{0.9}$ for $epoch = 1,...,N$, with $N=2000$ for the SAROS trained model and $N=4000$ for the model trained on the extended TotalSegmentator dataset. The weights with the best validation Dice Coefficient (DC) were used following training. Development was carried out on an NVIDIA GeForce RTX 4090 GPU.
\\
\textbf{Gaussian Kernel Density Estimation}\\
Following semantic segmentation, GKDE analysis is applied to the SAT predictions to refine the organ-free abdominal cavity mask and accurately segment VAT. 
Before fitting a GKDE to the SAT prediction, adjustments are necessary due to the inclusion of the cutis (the dermis and epidermis) in the SAROS ground-truth annotations. As the cutis is not AT, its intensity values do not accurately reflect those of SAT. Additionally, false positive predictions of SAT may predict voxels which are outside the body, further affecting the PDF estimated by the GKDE. To account for this, we erode the SAT prediction's outer layers at the \textit{L3} vertebral level, taken to be the median $z$ coordinate of the \textit{L3} prediction, until the mask is at $20\%$ of its original size. The GKDE is then fitted to the intensity values of this eroded mask. Fitting the GKDE to this axial prediction, as visualised in Fig.~\ref{fig:Fig3}, instead of the volumetric SAT prediction is chosen for computational efficiency. 

KDE methods estimate the underlying continuous probability density function (PDF) of a discrete dataset by smoothing out data points into overlapping continuous distributions, whose shape is determined by the kernel. 
%
In our application, we extracted all of the intensity values in the SAT segmentation pixels and formed a discrete set $S = \{{X_{1},...,X_{n}}\}$ with $X_{i} \in \mathbb{R}$ and $n$ the number of intensity values. To generate the entries of $x$ into the kernel, we uniformly sampled $1000$ points in the range $[I_{min},I_{max}]$, creating the set $\{x_{1},...,x_{1000}\}$, where $I_{min},I_{max}$ are the minimum and maximum HU intensity values of the predicted SAT. Thus for each $x_{j},j=1, \dots, 1000$, a kernel is estimated. In our application, we estimated a Gaussian kernel, $K$. Iterating over the $1000$ samples, we generate a GKDE:
\begin{equation}\label{full_gaussian_kde}
    p(x) = \frac{1}{nh}\sum^{1000}_{j=1}\sum^{n}_{i=1}K\left(\frac{x_{j}-X_{i}}{h}\right), \quad K\left(\frac{x-X_{i}}{h}\right) = \frac{\exp\left(-\frac{1}{2} \lVert\frac{ x - X_{i}}{h}\rVert^{2}\right)}{\int_{\mathbb{R}} \exp\left(-\frac{1}{2}\lVert\frac{ x - X_{i}}{h}\rVert^{2}\right) dx},
\end{equation}
where $\lVert \cdot \rVert$ is the Euclidean norm and $h$ is some smoothing factor taken to be Scott's factor $h = n^{-\frac{1}{5}}$ as it scales to the size of the dataset.
This GKDE gives a continuous approximation for the distribution of HU intensity values in SAT. 
We used Eq.~\ref{full_gaussian_kde} to refine the prediction of VAT voxels within the abdominal cavity mask by evaluating $p$ on their intensity values. The PDF $p$ maps the set of voxels in the abdominal cavity mask $\{y_{1},...,y_{m}\}$  to a set of probability densities (PDs) $\{p(y_{1}),...,p(y_{m})\}$. The PD value associated to each voxel intensity indicates how well it fits into the approximated continuous PDF, with higher values indicating a better fit. By removing a percentage of voxels with the lowest PD from the abdominal cavity prediction, the ones least likely to be VAT,  we obtain the final VAT prediction.
The full \textit{KEVS} deployment utilising semantic segmentation, SAT erosion, GKDE fitting, and VAT mask prediction is visualised in Fig. \ref{fig:Fig1}. The PD percentile used in \textit{KEVS} was $15\%$, chosen as this threshold best balances metrics performance, as shown in Online Resource 1 Fig. 1.

\section{Experiments and Results}\label{sec3}
We assessed the semantic segmentation performance of \textit{KEVS} in five different segmentation tasks across three datasets.
This includes the testing sets of the SAROS and TotalSegmentator datasets and the validation set of the publicly available AMOS22 dataset \cite{ji2022amos}. Additionally, we assessed the ability of \textit{KEVS} to accurately predict the abdominal organs separately as a binary label for the TotalSegmentator and AMOS22 datasets. For testing on the SAROS dataset, we only assessed the model's ability to segment SAT and the abdominal cavity as these are most relevant to our pipeline.

We compared \textit{\textit{KEVS}'s} ability for VAT segmentation against five traditional thresholding techniques and the TotalSegmentator model. This was carried out on the newly introduced UCLH-Cyst dataset.
We independently considered VAT predictions over the entire abdominal cavity and within the region bounded by the lumbar vertebrae. This restriction minimises false positives outside the abdominal area, providing additional clarity for evaluating model performance in the abdominal region. Following \cite{maier2024metrics}, we use the DC and Normalised Surface Distance with tolerance $\tau = 2$ (NSD) as our main performance metrics and also report Precision and Recall. 

\subsection{Validation Datasets}
Prior to prediction on both datasets, all images were resampled with trilinear resampling to have voxel spacing of (1.5, 1.5, 1.5), to align with the training of \textit{KEVS} and the TotalSegmentator model.

\textbf{AMOS22} contains 500 $3D$ CT scans with ground-truth annotation for 15 abdominal organs. This includes a pre-defined validation set of 100 CT scans. For our evaluation only this validation set was used, serving as an unseen validation set for the semantic segmentation models.

\textbf{UCLH-Cyst} consists of 20 CT scans performed before cystectomy procedures at UCLH, with ground-truth annotations for VAT. The ground-truth VAT masks were produced using a semi-automated method, whereby we use a segmentation technique to produce predictions which were then manually edited by three radiologists. To reduce bias towards a certain segmentation technique, ten were produced using the $(-190, -30)$ HU thresholding method, five using the TotalSegmentator tissue model, and five using \textit{KEVS}. 

\subsection{Semantic Segmentation Results}
Table~\ref{full_seg} presents segmentation results for \textit{KEVS} across three datasets and a variety of segmentation tasks. The obtained performance suggests that \textit{KEVS} is an accurate tool for CT segmentation. \textit{\textit{KEVS}'s} achieved very high accuracy in segmenting abdominal organs as a binary task, with a mean DC of $0.9588$ and $0.9217$ on the TotalSegmentator and AMOS22 datasets respectively. Evidently, accurate exclusion of abdominal organs for VAT consideration, as considered in our pipeline, is viable.
\begin{table}[t]
    \centering
    \resizebox{\textwidth}{!}{
    \begin{tabular}{ |c|c|c|c|c|c|c|  }
    \hline
    \multicolumn{1}{|c|}{} & \multicolumn{5}{c|}{\textbf{Metric}} \\
    \hline
    \textbf{Dataset} & \textit{DC} & \textit{NSD} & \textit{Precision} & \textit{Recall} & \textit{BDC}\\
    \hline
    \textit{TotalSegmentator (all original labels)}    &$0.8900	\pm 0.1052$ & $0.9113 \pm 0.1121$  & $0.9021 \pm 0.0944$ & $0.9262 \pm 0.0373$ & -\\
    \textit{TotalSegmentator  (abdominal organs)}    & $0.8686 \pm 0.0766$ & $0.8795 \pm 0.0886$  & $0.8767 \pm 0.0638$ & $0.9111 \pm 0.0595$ & $0.9588 \pm 0.0315$ \\
    \textit{SAROS (SAT)}    & $0.9656 \pm 0.0516$ & $0.9658 \pm 0.0506$  & $0.9634 \pm 0.0509$ & $0.9698 \pm 0.0482$ & -\\
    \textit{SAROS (Abdominal Cavity)}    & $0.9540 \pm 0.1294$ & $0.8796 \pm 0.1605$  & $0.9609 \pm 0.1114$ & $0.9655 \pm 0.0977$ & -\\
    \textit{AMOS22}  & $0.7938 \pm 0.1395$ & $0.7924 \pm 0.1547$  & $0.8482 \pm 0.0881$ & $0.7945 \pm 0.1453$ &  $0.9217 \pm 0.0350$\\
    \hline
    \end{tabular}
    }
    \caption{The results of U-Mamba segmentation of various body features in CT scans assessed for DC, NSD, Precision, Recall, Binary Dice Coefficient (BDC).}
    \label{full_seg}
\end{table}
\subsection{VAT Segmentation Results}
Table~\ref{vat_segmentation} lists VAT segmentation results for five HU thresholding techniques, the TotalSegmentator tissue model, and \textit{KEVS}, which outperforms all other methods across all metrics. When evaluating across the entire abdominal cavity (bounded region), \textit{KEVS} achieves a $4.80\%$ ($2.93\%$), $3.59\%$ ($3.50\%$), $3.76\%$ ($0.16\%$), and $0.96\%$ ($1.35\%$) increase in mean performance for DC, NSD, Precision, and Recall over the best non-\textit{KEVS} model respectively. 
\begin{table}[b]
    \centering
    \resizebox{\textwidth}{!}{
    \begin{tabular}{ |c|c|c|c|c|c|c|c|  }
    \hline
    \multicolumn{2}{|c|}{} & \multicolumn{4}{c|}{\textbf{Metric}} \\
    \hline
    \textbf{Body region} & \textbf{Method} & \textit{DC} & \textit{NSD} & \textit{Precision} & \textit{Recall} \\
    \hline
    \multirow{7}{*}{\parbox[c][7\baselineskip][c]{0cm}{\centering\begin{sideways}Full Cavity\end{sideways}}} & \textit{Thresholding (-190, -30)}    & $0.8203 \pm 0.1306$ & $0.8493 \pm 0.0995$  & $0.8180 \pm 0.0828$ & $0.8327 \pm 0.1662$\\
    & \textit{Thresholding (-195, -45)}  & $0.8043 \pm 0.1452$ & $0.8432 \pm 0.1169$  & $0.8401 \pm 0.0887$ & $0.7826 \pm 0.1748$ \\
    & \textit{Thresholding (-200, -10)}    & $0.8126 \pm 0.1067$ & $0.8215 \pm 0.0736$ & $0.7668 \pm 0.0787$ & $0.8724 \pm 0.1446$\\
    & \textit{Thresholding (-200, -20)}    & $0.8190 \pm 0.1185$ & $0.8395 \pm 0.0866$ & $0.7938 \pm 0.0800$ & $0.8547 \pm 0.1560$\\
    & \textit{Thresholding (-250, -50)}    & $0.7914 \pm 0.1504$ & $0.8303 \pm 0.1222$ & $0.8364 \pm 0.0978$ & $0.7616 \pm 0.1769$\\
    & \textit{TotalSegmentator}    & $0.8298 \pm 0.1182$ & $0.8311 \pm 0.0964$ & $0.8245 \pm 0.0629$ & $0.8485 \pm 0.1606$\\
    & \textit{\textit{KEVS} (ours)}    & $\mathbf{0.8697 \pm 0.0670}$  & $\mathbf{0.8798 \pm 0.0511}$ & $\mathbf{0.8679 \pm 0.1188}$ & $\mathbf{0.8808 \pm 0.0251}$\\
    \hline
    \multirow{7}{*}{\parbox[c][7\baselineskip][c]{0cm}{\centering\begin{sideways}Vertebral Bounds\end{sideways}}}& \textit{Thresholding (-190, -30)}    & $0.8414 \pm 0.1355$ & $0.8682 \pm 0.0992$  & $0.8444 \pm 0.0803$ & $0.8489 \pm 0.1705$\\
    & \textit{Thresholding (-195, -45)}  & $0.8264 \pm 0.1509$ & $0.8629 \pm 0.1173$  & $0.8623 \pm 0.0899$ & $0.8053 \pm 0.1806$ \\
    & \textit{Thresholding (-200, -10)}   & $0.8392 \pm 0.1113$ &	$0.8472 \pm 0.0752$ &	$0.8049 \pm 0.0726$ &	$0.8842 \pm 0.1485$\\
    & \textit{Thresholding (-200, -20)}    & $0.8325 \pm 0.1543$ &	$0.8630 \pm 0.1075$ &	$0.8197 \pm 0.0921$ &	$0.8581 \pm 0.1946$\\
    & \textit{Thresholding (-250, -50)}    & $0.8058 \pm 0.1927$ &	$0.8581 \pm 0.1535$ &	$0.8486 \pm 0.1270$ &	$0.7818 \pm 0.2163$\\
    & \textit{TotalSegmentator}    & $0.8743 \pm 0.1268$ &	$0.8804 \pm 0.1014$ &	$0.9065 \pm 0.0586$ &	$0.8600 \pm 0.1641$\\
    & \textit{\textit{KEVS} (ours)}    & $\mathbf{0.9000 \pm 0.0429}$  & $\mathbf{0.9112 \pm 0.0331}$ & $\mathbf{0.9080 \pm 0.0795}$ & $\mathbf{0.8961 \pm 0.0258}$\\
    \hline
    \end{tabular}
    }
    \caption{The results of seven different methods for segmentation of abdominal VAT in $3D$ CT scans, including five HU thresholding techniques, TotalSegmentator, and \textit{KEVS} (ours). We assess for DC, NSD, Precision, Recall. The best result as determined by mean value in each metric is highlighted in bold.}
    \label{vat_segmentation}
\end{table}

To evaluate statistical significance, we extracted the axial slices from the UCLH-Cyst scans and computed segmentation metrics for each slice. We then applied a one-sided Wilcoxon signed rank test to determine if the difference in the segmentation prediction of the different models was statistically significant, considered if $p<0.05$. When comparing \textit{KEVS} to TotalSegmentator we evaluated the slices of the ten scans annotated through semi-automation with HU thresholding, and when comparing \textit{KEVS} to thresholding we used the slices of the five scans annotated through semi-automation with TotalSegmentator. This is done to reduce bias which may occur due to an unbalanced number of axial slices being produced by each method due to differing scan sizes. To standardise the comparison between the two tests, we assessed the region bounded by the lumbar vertebrae. This produced 901 and 447 annotated axial slices respectively. In both tests, \textit{\textit{KEVS}'s} improvement in segmentation performance  was statistically significant over all other models for DC, NSD, and Precision.
\subsection{GKDE vs Thresholding on Organ Free Abdominal Cavity}
%
To assess the value of applying GKDE instead of manual HU thresholding to the organ free abdominal cavity mask to make VAT predictions, we also recorded the metrics for each thresholding technique applied to this mask. When comparing over the entire cavity (bounded region), the use of GKDE improves DC, NSD, and Recall by an average of $4.03\%$ ($4.90\%$), $0.98\%$ ($2.09\%$), and $11.95\%$ ($11.08\%$) respectively, but decreased Precision by $6.06\%$ ($4.30\%$). This suggests that using manual HU thresholding over GKDE causes VAT underprediciton.
%
\begin{figure}[b]
    \centering
    \includegraphics[width=1\textwidth]{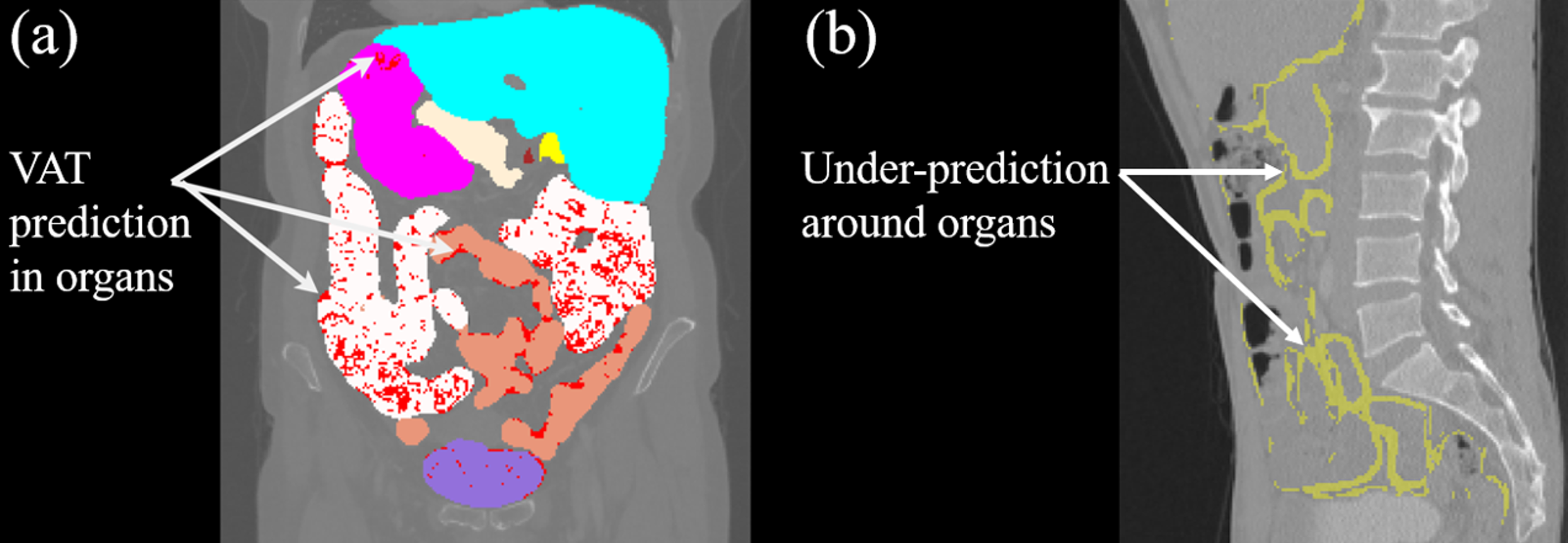}
    \caption{\textbf{(a)} Prediction of VAT inside abdominal organs for (-190, -30) thresholding.\textbf{(b)} An example of false-negative VAT prediction (yellow) around abdominal organs for TotalSegmentator model. In both cases, these erroneous predictions are reasons for reduction in predictive performance when compared to \textit{KEVS}.}
    \label{fig:Fig4}
\end{figure}
\section{Discussion and Conclusion}\label{sec4}
\begin{figure}[b]
    \centering
    \includegraphics[width=1\textwidth]{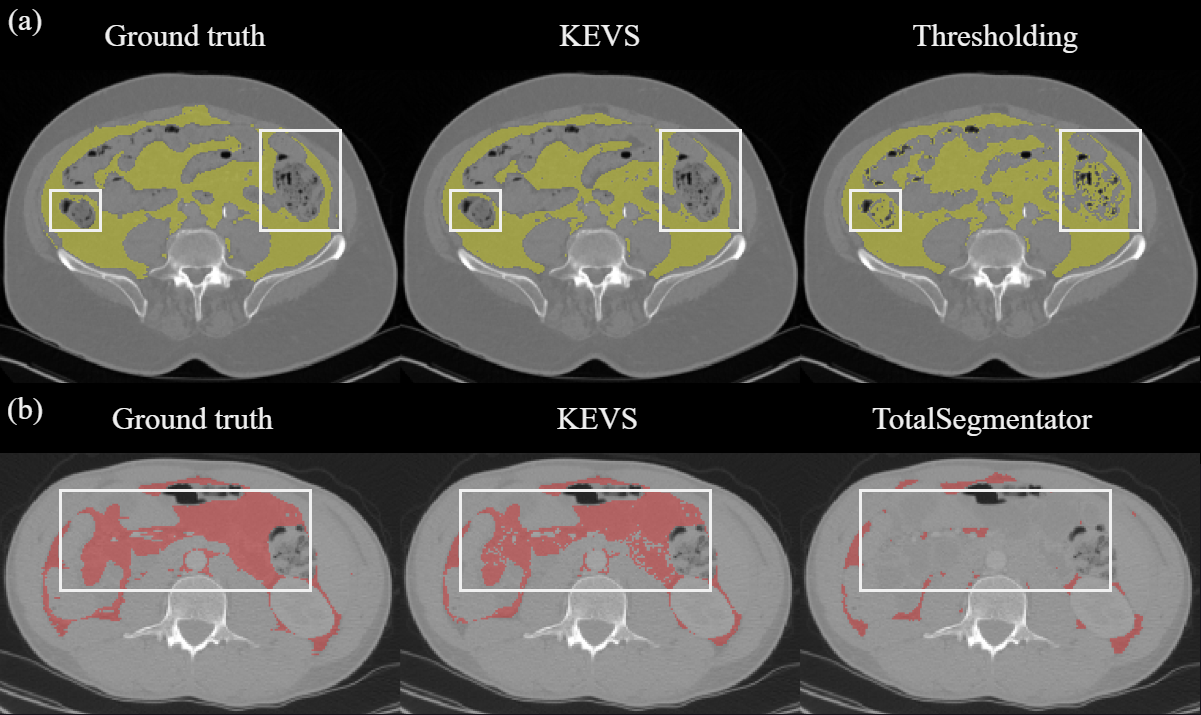}
    \caption{\textbf{(a)} A comparison of ground-truth mask, \textit{KEVS}, and TotalSegmentator prediction on an axial slice \textbf{(b)} A comparison of ground-truth mask, \textit{KEVS}, and (-190,-30) thresholding prediction on an axial slice. In both cases, the white boxes outline examples of \textit{KEVS} better performance in comparison to the alternative method.}
    \label{fig:Fig5}
\end{figure}

We have demonstrated that for VAT segmentation, \textit{KEVS} outperforms, across all metrics, traditional HU thresholding and the TotalSegmentator model. Improvement over thresholding techniques is due to two reasons. First, HU thresholding is susceptible to predicting organ tissue as VAT. In the UCLH-Cyst dataset, an average of $11.70\%$ of the HU thresholding VAT prediction coincides with abdominal organs. An example of this is illustrated in Fig. \ref{fig:Fig4}(a), and in Online Resource 2. By construction, \textit{KEVS} makes no such erroneous predictions. A comparison between a ground-truth mask, \textit{KEVS} prediction, and thresholding prediction showcasing this disparity is displayed in Fig.~\ref{fig:Fig5}(a). Secondly, the ability of \textit{KEVS} to model the scan-specific AT distribution provides it with additional context to produce accurate VAT predictions, which manual thresholding does not allow for. This accounts for the sustained difference in performance when simply applying an HU threshold to the organ free abdominal cavity.
The improved performance of \textit{KEVS} over TotalSegmentator is attributed to the latter model's increased tendency to under-predict VAT, particularly around abdominal organs. When considering only the first two voxels layers around the abdominal organs on the ten scans annotated through semi-automation with HU thresholding, we record a $10.88\%$ decrease in DC from \textit{KEVS} to TotalSegmentator; highlighting this disparity. This is reflected in Table~\ref{vat_segmentation} with the comparatively lower Recall score, but a more similar Precision score compared to \textit{KEVS}. An example of this under-prediction is shown in Fig.~\ref{fig:Fig4}(b) and when compared to ground-truth and \textit{KEVS} in Fig.~\ref{fig:Fig5}(b).

The use of GKDE in addition to semantic segmentation in \textit{KEVS} increased the average VAT prediction time on a scan by $72.08$s over TotalSegmentator ($20.26$s).  We believe the potential downsides of this time increase are outweighed by the increased prediction performance and does not prevent clinical deployment of \textit{KEVS}. 

This paper introduces \textit{KEVS}, a novel method for VAT segmentation in pre-cystectomy CT scans combining DL segmentation models and GKDE modelling of AT. Notably, \textit{KEVS} does not require ground-truth VAT masks during training and adapts to the specific characteristics of each CT scan. Future work will evaluate its generalisability to peri-operative CT images of general cancer surgery. As suggested by the results in Online Resource 1 Table 1, incorporating \textit{KEVS} analysis could enhance predictions of post-cystectomy complications beyond the use of pre-operative patient characteristics alone, where the inclusion of \textit{KEVS} body tissue analysis information improves the ability of machine learning models to predict complications in $73\%$ of evaluations. This highlights the potential for broader use of \textit{KEVS} for comprehensive body tissue analysis prior to general cancer surgery. 
Future research will also focus on dynamically adjusting the PD threshold to optimise VAT segmentation performance.

Overall, \textit{KEVS} presents a novel, SOTA method for the analysis of VAT in pre-cystectomy CT scans. \textit{KEVS} is fully automated, eliminating inter-observer variability, while GKDE modelling enables effective adaption to any-dose CT scan. An example is provided in the video in Online Resource 2 where \textit{KEVS} adapts more effectively for VAT prediction on a noisy scan compared to TotalSegmentator. Owing to its flexibility, \textit{KEVS} has significant potential for clinical application in peri-operative CT analysis towards prediction of post-surgical outcomes. 

%
\backmatter
\section*{Statements and Declarations}
\bmhead{Funding}
This work was supported by the EPSRC-funded UCL Centre for Doctoral Training in Intelligent, Integrated Imaging in Healthcare (i4health) [EP/S021930/1]; the International Anesthesia Research Society Mentored Research grant, and the NIHR University College London Hospitals Biomedical Research Centre.
\bmhead{Conflict of interest} 
The authors declare no conflict of interest.
\bmhead{Ethics approval} 
Local Research Ethics Committee approval was given for the collection of patient data on an institutional database.
\bmhead{Informed consent} 
Patients provided written informed consent for review of their clinical data (IRAS ID: 255541, REC Reference: 19/LO/1371).
\bmhead{Code, data and materials availability} 
Code and the full dataset will be made publicly available in our institutional database (www.rdr.ucl.ac.uk), upon paper acceptance, for research purposes.

\bibliography{sn-article}

\end{document}